\documentclass[reprint, pra,twocolumn,superscriptaddress,floatfix,amsmath,amssymb]{revtex4-1}

\usepackage[T1]{fontenc}
\usepackage{float}
\usepackage{graphicx}
\usepackage{bm}
\usepackage{braket}
\usepackage{xcolor}
\usepackage{amsfonts}
\usepackage{comment}
\usepackage{dirtytalk}
\usepackage{url}
\usepackage{bbold}
\usepackage{realboxes}

\usepackage{enumitem}
\usepackage[toc,page,titletoc]{appendix}

\usepackage{tocloft}
\definecolor{dark-red}{rgb}{0.4,0.15,0.15}
\definecolor{dark-blue}{rgb}{0.15,0.15,0.4}
\definecolor{medium-blue}{rgb}{0,0,0.5}
\usepackage[hypertexnames=false]{hyperref}
\usepackage[all]{hypcap}
\hypersetup{
    colorlinks, linkcolor={dark-red},
    citecolor={dark-blue}, urlcolor={medium-blue}
}
\usepackage{cleveref}

\usepackage{algpseudocode}

\newcommand{\cmmnt}[1]{}

\usepackage{listings}
\usepackage{xcolor}

\definecolor{bkgd}{RGB}{240,242,246}
\definecolor{ceruleanblue}{rgb}{0.16, 0.32, 0.75}
\definecolor{orange-red}{rgb}{1.0, 0.27, 0.0}
\definecolor{anotherblue}{RGB}{37,92,243}
\definecolor{blackblue}{RGB}{46,60,85}
\definecolor{goldyellow}{RGB}{199,146,12}
\lstdefinestyle{altstyle2}{
    backgroundcolor=\color{bkgd},
    basicstyle=\ttfamily\footnotesize\color{blackblue},
    breakatwhitespace=false,
    breaklines=true,
    captionpos=b,
    commentstyle=\color{goldyellow},
    keepspaces=true,
    keywordstyle=\color{orange-red},
    language=Python,
    numbersep=5pt,
    numberstyle=\tiny\color{ceruleanblue},
    showspaces=false,
    showstringspaces=false,
    showtabs=false,
    stringstyle=\color{anotherblue},
    tabsize=2
}
\begin{document}

\title{Noisy dynamical systems evolve error correcting codes and modularity
}

\author{Trevor McCourt } \thanks{tmccourt@mit.edu}
\affiliation{Department of Electrical Engineering and Computer Science, Massachusetts Institute of Technology, Cambridge, MA 02139, USA
}\affiliation{The NSF AI Institute for Artificial Intelligence and Fundamental Interactions, Cambridge, Massachusetts 02139, USA}

\author{Ila R. Fiete}
\affiliation{Department of Brain and Cognitive Sciences, MIT, Cambridge, MA, USA
}
\affiliation{McGovern Institute for Brain Research, Department of Brain and Cognitive Sciences, Massachusetts Institute of Technology, Cambridge, MA 02139
}

\author{Isaac L. Chuang}
\affiliation{Department of Electrical Engineering and Computer Science, Massachusetts Institute of Technology, Cambridge, MA 02139, USA
}
\affiliation{Department of Physics, Massachusetts Institute of Technology, Cambridge, MA 02139, USA
}
\affiliation{The NSF AI Institute for Artificial Intelligence and Fundamental Interactions, Cambridge, Massachusetts 02139, USA}

\date{\today}

\begin{abstract}

Noise is a ubiquitous feature of the physical world. As a result, the first prerequisite of life is fault tolerance: maintaining integrity of state despite external bombardment. Recent experimental advances have revealed that biological systems achieve fault tolerance by implementing mathematically intricate error-correcting codes and by organizing in a modular fashion that physically separates functionally distinct subsystems. These elaborate structures represent a vanishing volume in the massive genetic configuration space. How is it possible that the primitive process of evolution, by which all biological systems evolved, achieved such unusual results?  In this work, through experiments in Boolean networks, we show that the simultaneous presence of error correction and modularity in biological systems is no coincidence. Rather, it is a typical co-occurrence in noisy dynamic systems undergoing evolution. From this, we deduce the principle of error correction enhanced evolvability: systems possessing error-correcting codes are more effectively improved by evolution than those without.

\end{abstract}

\maketitle

Even under the influence of entirely deterministic natural laws, a finite agent that interacts with a large environment will experience some degree of randomness. This is because it will eventually lose record of all its previous interactions, and will only be able to reason on probabilistic terms. This is exemplified in physics by observations of Brownian motion \cite{einstein1905motion}, in which an observed particle in contact with a large bath of unobserved particles undergoes seemingly random, yet entirely determined, motion. Therefore it is reasonable to suggest that anywhere life emerges, it is probably doing so despite a stochastic environment, as it has on Earth. 

Von Neumann appreciated the fundamental nature of noise and observed that the biological information processing systems that underpin life handle it elegantly \cite{vonneumann51}. In particular, biological systems are generally not halted by a single component failure or error. In contrast, the engineered computers of his time were generally brittle, and errors in a single component had to be identified and addressed before the rest of the system could continue operation. Von Neumann thought this approach would begin to become infeasible as more and more complex systems were constructed \cite{vonneumann66}. To resolve this and to contribute further to our understanding of the connection between computing in the presence of noise and life, he established the theory of fault-tolerant computation \cite{vonneumann56}.

Fault-tolerant computing systems compose noisy primitive computational units (e.g. logic gates) into structures that compute the same primitive, robustly \cite{vonneumann56, ftintro}. This concept is not limited to digital computation: fault tolerance has recently been demonstrated in neural networks with brain-like representations \cite{Zlokapa2022} and is also an active area of theoretical and experimental research in quantum computing \cite{shor_ft, surfacecode, Andersen_2020, Krinner_2022}, where it is particularly important due to the high susceptibility of real qubits to noise. Today, fault-tolerant computing is generally accomplished by computing within the context of an error-correcting code. An error-correcting code specifies a way to robustly store or transmit messages in the presence of noise \cite{shannon48}. The basic idea is to encode messages in codewords, which adds redundancy such that if part of the codeword is lost the message can still be recovered. Fault-tolerant computing further requires that the recovery be done by the noisy system itself. 

Since the time of von Neumann, many specific examples of robust behavior in biological information processing systems have been discovered. Genomes are one example of this. Single-knockout experiments in a particular yeast found that only $19 \%$ of the genes were required for the organism to continue living \cite{Dow2002, Deutscher2006}. The same resilience is seen in the human brain, which can tolerate substantial death of neurons with minimal cognitive impairment \cite{Drachman2005, Mosley2005}. This level of robustness implies the existence of mechanisms for error correction. Studying these mechanisms is an emerging topic of theoretical interest in genetics \cite{Faria2012}. There is also substantial evidence for formal error correction mechanisms in the brain. Many neural circuits have population activities that live on low dimensional manifolds \cite{Yoon2013,Chaudhuri2019, Gardner2022, Khona2021AttractorAI}, reminiscent of topological error-correcting codes \cite{nat_top, lidar_brun_2013}. In particular, it is thought that grid cells use an efficient error-correcting code to robustly encode position in the presence of noise \cite{Sreenivasan2011}.

One way that large-scale natural and engineered systems seem to achieve robustness is through {\em modularity}, the division of a large system into functionally independent parts. The genome is believed to have spatial and functional modularity \cite{Zheng2022}, and a common view of the brain is that it is composed of many modules responsible for different functions that combine to form our intelligence \cite{Meunier2010}.
At a large scale, organizations, processes, and machines frequently have a division of responsibility for separate subtasks, such that if one subsystem fails the system can still complete its function \cite{Landau1969, lerner86, nonaka90}.  In general, an overarching motif observed in natural systems is that they appear to be assembled out of smaller, individually robust components.

At the same time, biological structures emerge by making local changes that improve fitness, a process of adaptive evolution.  It is incredible that evolution was able to produce modular, error-correcting structures through this process, given that they represent a vanishing fraction of the configuration space of possible solutions to a given problem. A tantalizing question is therefore what conditions and principles drive evolution toward such structures? The answer to this question is significant, as it gets to the root of forces that drive the crystalization of life.

Any property that increases the efficacy of adaptive evolution may be referred to as evolvable. Evolvable properties should emerge as a typical result of evolution since obtaining some of an evolvable property increases an organism's ability to acquire more of it. Identifying evolvable properties is a topic of constant interest \cite{Arndt2000, Pigliucci2008, Payne2019, Vaishnav2022}.  For example, Kauffman famously showed that systems with a phase transition from frozen to chaotic behavior (a so-called "ordered" phase) tend to be more evolvable, and seek the edge of chaos \cite{Kauffman1992}. 

\begin{figure}
    \centering
    \includegraphics[width=\linewidth]{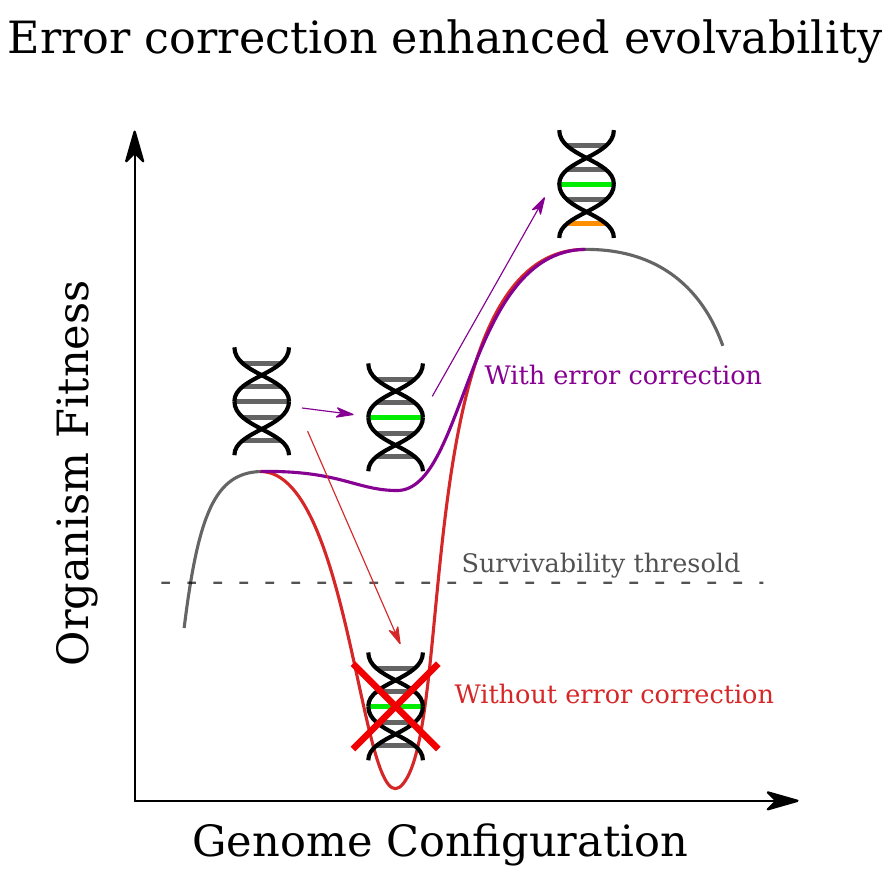}
    \caption{\textbf{Error correction smooths valleys in fitness landscapes. }Noisy systems initially evolve error correction to protect themselves from random environmental fluctuations that affect their state. However, once evolved, error correction also helps \textit{neutralize} permanent genetic mutations that would otherwise be lethal. During evolution, this allows organisms that possess error-correcting codes to search further in genome configuration space for improvements than their non-error-correcting counterparts. This is illustrated in the figure: the organism that uses error correction can sustain the first genome mutation without falling into a valley of fitness that is too low to survive. It can then acquire a further mutation which brings it to a new, improved local fitness maximum, corresponding to an improved error-correcting code. With this new code, it will even more easily be able to find improvements, and so on. Error correction begets more error correction, leading to its ubiquity in the natural world.}
    \label{fig:intro_fig}
\end{figure}

In this work, we show that both error-correcting codes and modularity are typical co-emergent results of evolution in a noisy environment. We show that this occurs because organisms with error-correcting properties are better protected from lethal mutations than those without, allowing them to more effectively search configuration space for improvements. From this, we introduce the concept of error correction-enhanced evolvability: organisms with error-correcting codes are more evolvable than those without. This principle is illustrated in Fig.~\ref{fig:intro_fig}. Noise bootstraps this phenomenon, suggesting that noise plays an important role in evolving complex structures. 

We begin by introducing Boolean networks and motivating their use in the study of evolution and life. We then show that when evolved to perform primitive computations in the presence of noise, Boolean networks almost always develop strong error-correcting codes. By probing the effect of mutations on evolved networks, we then demonstrate that this occurs because error-correcting codes increase the number of neutral mutations seen by organisms, allowing them to more effectively search locally for improvements than their non-error-correcting counterparts. Finally, we show that under composite computational tasks (that require more than one primitive), evolution tends to produce structurally modular organisms that employ modular error-correcting codes, beautifully reflecting what is seen in biology.

Originally developed to model gene regulatory networks \cite{kauffman69, Kauffman1969HomeostasisAD, THOMAS1973563}, Boolean networks are autonomous, discrete-time dynamical systems with binary-valued states. Boolean networks can be viewed as dynamic function evaluators. A function input can be given as an initial state to a network and the output is simply the state at some specified later time. Boolean networks are therefore reasonable computational models of primitive living organisms: receiving input and later producing the correct output is equivalent to an agent responding to environmental stimulus to avoid death \cite{active_inf}. Despite their simple binary nature, Boolean networks can express very complex behavior. For example, they possess the aforementioned ordered phase that seems to be crucial for successful adaptation \cite{Kauffman91, Kauffman1992}. This complexity makes them a candidate for modeling large-scale brain dynamics \cite{Bertacchini2022}. 

\begin{figure*}
    \centering
    \includegraphics[width=\textwidth]{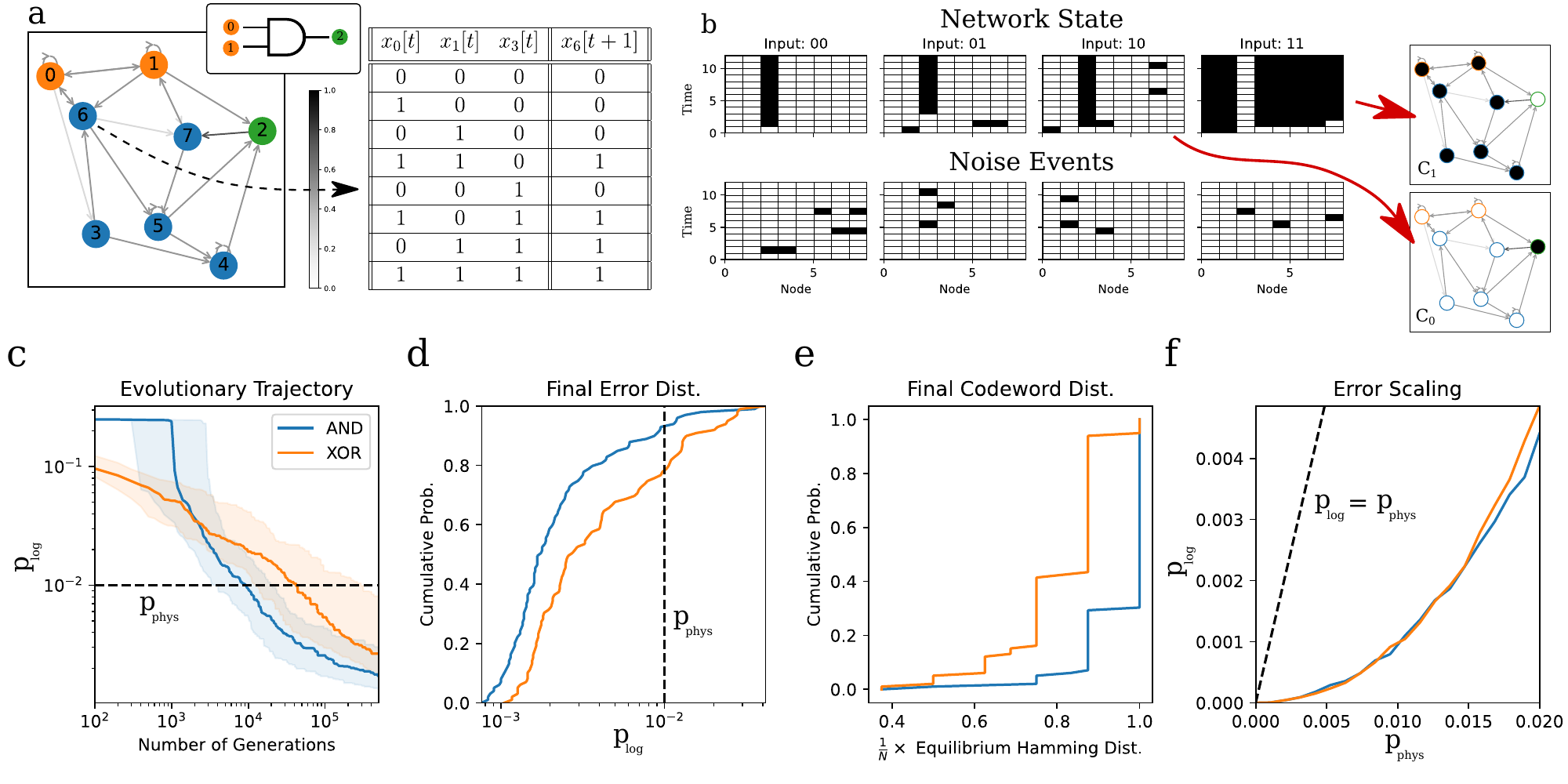}
    \caption{\textbf{Boolean networks adapting into error correction. a} A Boolean network that has adapted to solve the AND task in the presence of noise. Inputs are provided to the orange nodes 0 and 1 at time t=0 and, the answer is expected at the green node 2 at t=T. A Boolean network is specified by a directed graph with a truth table living on each node. The state of each head node at time t+1 is a binary function of the state of each tail node at time t. For example, the state of node 6 at time t+1 is computed by applying a length $2^3$ truth table to the states of nodes 0, 1, and 3 at time t. The edge coloring reflects the influence of a head node on a tail node. Influence is the probability (taken over the entire truth table) that the head node flips if the tail node flips. \textbf{b} Examples of noisy dynamical trajectories of the network in (a) for each of the $2^2$ possible input states. The network encodes the answer in codewords of maximal Hamming distance and stabilizes the codeword against noise events. Codewords may be taken as the final state of the network $x[T]$ for the different output values. The two codewords are visualized on the network graph. The boundary of the circles indicates the node function, and the fill indicates the codeword. \textbf{c} Evolutionary trajectories of 150 randomly initialized populations learning to solve the XOR and AND tasks. Populations were not pruned, every population that was started is included in the statistics. The solid line indicates the median logical error probability, and the shaded region indicates the interquartile range.  \textbf{d} Distribution of final logical error probabilities over all populations. In both tasks, typical organisms learn to suppress errors far below the physical noise level ($p_{\text{phys}}$, indicated by the dashed black line). \textbf{e} Distribution of average Hamming distance between the codewords $C_0$ and $C_1$ over all populations. The typical organism uses a strong error-correcting code to suppress errors. \textbf{f} Scaling of the logical error probability with the physical error probability for the highest performing organisms: logical error probability is strongly suppressed relative to physical error probability. }
    \label{fig:emergent_coding}
\end{figure*}

Here, we leverage Boolean networks to computationally probe evolution in a noisy environment. We assign Boolean network organisms a measure of fitness based on their ability to compute a simple function in the presence of noise. We provide an organism with the input to the computation in a raw, unencoded form, and allow it to undergo noisy dynamics for several timesteps. It is then queried for the output, which can be compared to the correct output for fitness calculation. Producing the correct output in the presence of noise is a daunting task: an organism has to have a method for encoding the inputs, robustly performing the computation, and stabilizing the answer against noise in memory. We study the evolution of organisms that compute primitive and more complicated composite computations to study the emergence of error correction and modularity.

The state of a Boolean network of $N$ nodes is given by a length $N$ binary vector $x$. The state of the $i^{th}$ node $x_{i}$ at time $t+1$ is given as an arbitrary Boolean function $f_i$ of the state of $k_i \leq k_{max}$ other nodes at time $t$. Therefore, a Boolean network can be thought of as a directed graph with truth tables of length $2^{k_i}$ living on each node (Fig.~\ref{fig:emergent_coding}{a}) The genome of such a Boolean network is therefore specified by a matrix containing the truth tables for each node and a matrix defining the connectivity between nodes. For further information on how the Boolean network genome was specified in these experiments, see \cite{McCourt2023} or supplementary section 1.

The space of possible Boolean networks is truly massive. For a network with exactly $k$ connections per node, the number of unique networks is \cite{Gershenson2004, Harvey1997}
\begin{equation}\label{eqn:network_counting}
    \left( \frac{2^{2^k} N!}{(N-k)!}\right)^N
    \,.
\end{equation}
The networks considered in this work are {\em ragged}, which means that they have $k_i \leq k_{max}$ connections per node. As such, the number of networks is larger than what is indicated by Eq.~\ref{eqn:network_counting}. For the $N=8$ and $k_{max}=3$ network shown in Fig.~\ref{fig:emergent_coding}{a}, this evaluates to $\sim 10^{40}$. This number grows very rapidly with $N$; for $N=15$ there are more configurations than there are bits of information in the observable universe \cite{Vopson2021}. Therefore, any particular configuration of a Boolean network resides in a vanishing volume of configuration space. At the same time, Boolean network loss landscapes are generally rugged and multi-peaked, making adaptation non-trivial \cite{Kauffman1992}. Therefore, any configuration that is consistently reached by randomly initialized adaptive walks must be somehow attractive: it is exceptionally difficult to stumble upon particular solutions by accident when working with Boolean networks. This makes Boolean networks particularly useful for studying evolvability.

We introduce stochasticity to our Boolean networks \cite{Shmulevich2002} by injecting independent Bernoulli noise alongside the deterministic state update described above:
\begin{equation}\label{eqn:state_update}
    x[t+1] = f(x[t]) \oplus \text{Bernoulli}(p_{\text{phys}})
    \,,
\end{equation}
where $f$ is the collection of all the $f_i$ and represents the deterministic update previously described, $\oplus$ indicates bitwise XOR (exclusive OR), and $p_{phys} \sim 10^{-2}$ is the physical error probability. In other words, the noise acts to flip each bit of $f(x[t])$ independently with probability $p_{\text{phys}}$. Fig~\ref{fig:emergent_coding}{b} shows an example of the state of a noisy Boolean network vs time for different inputs.

We provide input to our organisms via designated input nodes, which are initialized with the input to a computational problem at $t=0$. For example, an organism that solves the AND task has two input nodes, as shown in Fig~\ref{fig:emergent_coding}{a}, which could be initialized as any two-bit binary string. The rest of the nodes are initialized to 0. The organism then undergoes $T-1$ steps of noisy dynamics, corresponding to deterministic state updates followed by exposure to noise (Eq.~\ref{eqn:state_update}). One more state update is then applied, serving as a noiseless decoding step, as standard in the study of fault-tolerant subsystems \cite{vonneumann56}. The state of the output nodes is then taken as the logical output of the organism. We randomize $T$ within some small range to force the organism to stabilize the solution instead of being allowed to pass through it transiently. The structure and complexity of the computational task can be varied to probe different properties of the solutions. The unfitness (or \emph{loss}) of a given organism is its logical error probability $p_{\text{log}}$. $p_{\text{log}}$ is the probability that the output node contains the wrong value at time T (after the decoding step), averaged over noise and all possible input binary strings. As an example, for the network shown in Fig~\ref{fig:emergent_coding}{a}, $p_{\text{log}}$ is the probability that at time $T$ the output node does not contain the AND of the values initially supplied to the input nodes. $p_{\text{log}}$ therefore reflects the effect of physical error $p_{phys}$ on the network.  For a fixed $p_{phys}$, a low $p_{\text{log}}$ means the network is performing the computation in a way that is robust to noise.

Endowed with a measure of fitness, we can then experiment with adaptive evolution in configuration space. Starting from a random configuration, organisms of increasing fitness are found by manipulating the genome via pairwise crossover operations (breeding) and local mutations. This procedure is intentionally primal and is meant to mirror operations that would have been feasible during bio-genesis. We tried a few different crossover operators, including no crossover (pure local search), and we found none of them substantially outperformed the local search (supplementary section 4).

Despite the apparent simplicity of Boolean networks, exploring their evolution in the presence of noise is extremely computationally intensive. For every evolutionary step, several hundred populations of several hundred organisms must be run through several hundred different noise trajectories starting from tens of different input states. This number quickly approaches batches of $10^8$ networks. Such large experiments were made possible by accelerating evaluation using modern Graphics Processing Units (GPUs) with large memories  \cite{McCourt2023}. Trillions of networks were simulated in the course of this work, consuming several GPU months of compute time.

We performed several experiments that probed the ability of Boolean networks to robustly perform the primitive computations AND and XOR in the presence of noise. Fig~\ref{fig:emergent_coding} a shows a typical resulting organism for the AND task (XOR results are similar, see supplementary section 3). The organism itself has little structure, as expected for a primitive task. One interesting point is that all of the resulting Boolean functions are of moderate influence. This is expected for a fault-tolerant solution, as heavy influence edges more readily propagate errors. What is most interesting about the evolved organisms is that they seem to achieve low error rates by implementing error-correcting codes. There are two possible answers to both the AND and XOR tasks: 0 and 1. As shown in Fig~\ref{fig:emergent_coding}{b}, the network given in Fig~\ref{fig:emergent_coding}{a} rapidly converges to a fixed point corresponding to the correct output for a given input, and the equilibrium state in each case may be taken as a codeword $C_0$ or $C_1$. The codewords are shown overlayed on the network graphs in Fig~\ref{fig:emergent_coding}{c}. The codewords are of maximum Hamming distance from each other, meaning that $C_0$ differs from $C_1$ by every possible bit. This is desirable, as it means the two states are maximally distinguishable in the presence of noise. The evolved organism also stabilizes the codeword: the state returns to the codeword one or two timesteps after perturbation by noise. 

\begin{figure*}
    \centering
    \includegraphics[width=\textwidth]{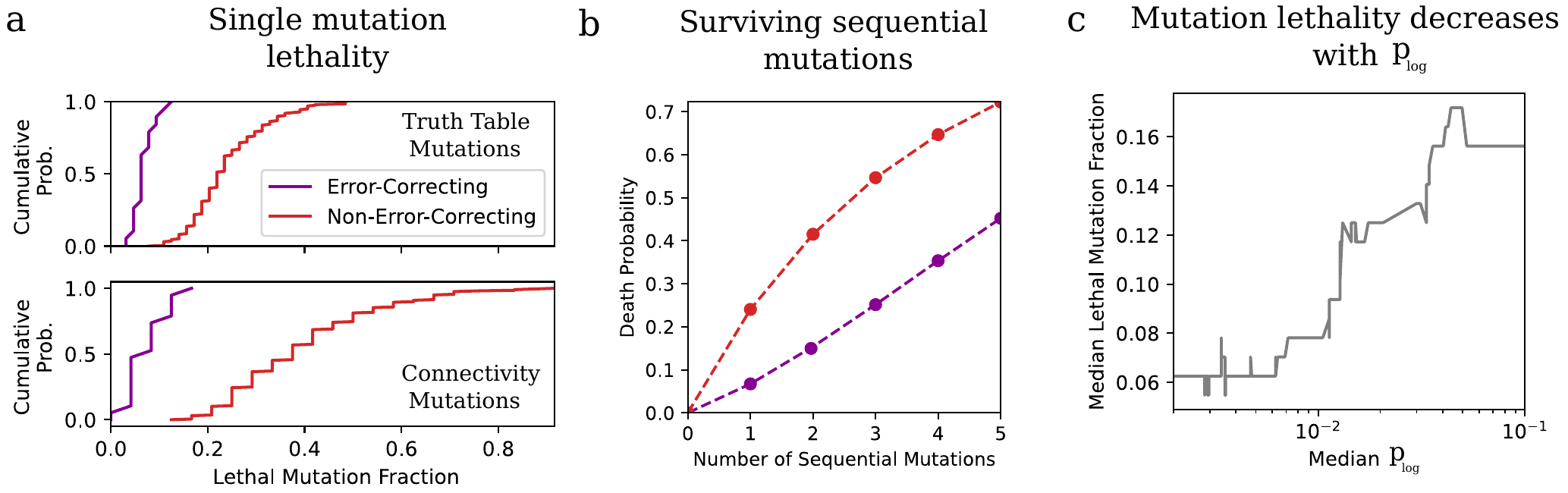}
    \caption{\textbf{The effect of error correction on evolvability. a} The lethality of different kinds of mutations for error-correcting and non-error-correcting AND organisms. A mutation is "lethal" if an organism that computes AND perfectly in the absence of noise can no longer do so after the mutation. Mutations to the truth tables consist of all possible bitflips applied to each node's truth tables. Mutations to the connectivity constitute adding or deleting a single connection between nodes at random. The lethal mutation fraction is the fraction of all of the possible mutations that lead to death for a given organism. This is plotted as a distribution over all organisms for each error type. \textbf{b} Comparing the probability of surviving several sequential truth table mutations for error-correcting and non-error-correcting AND organisms. Error-correcting organisms can survive three sequential mutations with the same probability that a non-error-correcting organism can survive one.  \textbf{c} Comparing the median logical error probability to the fraction of lethal truth table mutations for the populations of AND organisms from Fig.~\ref{fig:emergent_coding}. We see that the lethal mutation fraction decreases with logical error, a direct demonstration of the principle of error correction enhanced evolvability.  }
    \label{fig:coding_exp}
\end{figure*}

What is remarkable is that error-correcting codes seem to be a typical result of this experiment. The lower half of Fig.~\ref{fig:emergent_coding} shows results from 150 independent evolutionary experiments. These populations were all completely randomly initialized, and the results were not pruned: every population that was initialized has been included in the statistics. The distribution of population fitnesses over time is summarized in Fig.~\ref{fig:emergent_coding}{c}. As can be seen in Fig.~\ref{fig:emergent_coding}{d}, over $80 \%$ of the organisms compute the primitive with $p_{\text{log}} < p_{\text{phys}}$, with the median organism suppressing error by nearly an order of magnitude. Fig.~\ref{fig:emergent_coding}{e} shows that organisms accomplish this using error-correcting codes. Most organisms develop codewords of nearly maximum Hamming distance. Fig.~\ref{fig:emergent_coding}{f} shows that the organism suppresses logical error super-linearly, a hallmark of error correction.

It should be emphasized here that no prior information about error-correcting codes was supplied to the evolutionary procedure in the form of extra loss terms or otherwise: the mathematical concept of error correction emerges here exclusively from the primitive concept of surviving a noisy environment. This separates this result from previous studies where tailored optimization procedures were used explicitly to find good error-correcting codes \cite{popar_code_ev, ecsearch, ec_gen}.

The overwhelming typicality of error-correcting codes in these results is surprising. Complete convergence of so many randomly initialized populations is not expected for adaptive walks on rugged loss landscapes \cite{Kauffman1992}. We would expect many populations to get trapped in local minima and not find good codes. The implication of all of this in the presence of noise evolutionary processes somehow seek error correction. Note that the search of configuration space performed by evolution was nowhere near exhaustive. From Fig.~\ref{fig:emergent_coding}{c}, we can see that convergence was achieved in $\sim\! 10^6$ generations, with $\sim\! 10^2$ variations being explored each generation. The adaptive procedure explored a minuscule fraction of configuration space, $10^8/10^{40} = 10^{-32}$. This is far beyond finding a needle in a haystack ($\sim\! 10^{-10}$) and is closer to looking in the same haystack for a particular carbon atom.

One explanation is that error-correcting codes not only protect an organism from noise that strikes randomly during the dynamics but also from \emph{systematic} manipulation of the genome.  This is supported by data presented in Fig.~\ref{fig:coding_exp}. For non-error-correcting organisms, most genome mutations are lethal, such that an organism that perfectly performs a particular computation in the absence of noise before mutation no longer does so after mutation. However, as shown in Fig.~\ref{fig:coding_exp}{a}, the same is not true for error-correcting organisms, for which only a small percentage of mutations are lethal.

\begin{figure*}
    \centering
    \includegraphics[width=\textwidth]{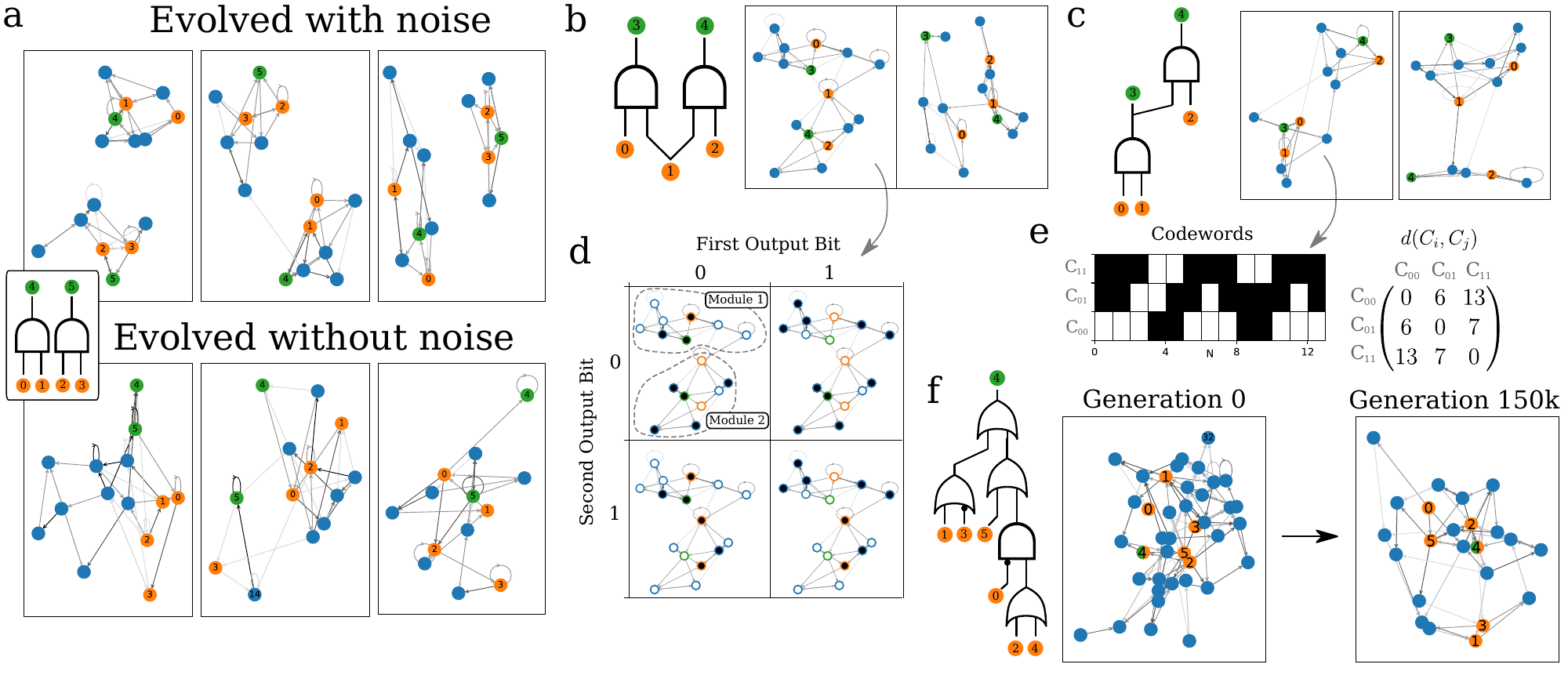}
    \caption{\textbf{Emergent structure in noisy adapting dynamic systems. a} Organisms that have adapted to solve the parallel AND task with and without noise, 
 all other experimental parameters held equal. The structures of the noisy organisms reflect the structure of the computational task, while the noiseless organisms have no obvious structure. This implies that noise alone can lead to modular dynamic systems. Experiments were also completed using the shared (\textbf{b}) and sequential (\textbf{c}) AND tasks. The computational structure of these tasks is also reflected beautifully in the evolved organisms. \textbf{d} An example of the codewords developed for the shared AND task. The organism possesses two distinct memory modules of equal size, each of which implements an independent maximum Hamming distance code. \textbf{e} The codewords associated with the shared AND task, which has three possible outputs. The codewords are all of maximum Hamming distance from one another.  \textbf{f} The structure of a network that computes a six-bit function before and after adaptation. We can see that the adaptive process imprints the computational structure on the organism, and also leads to the pruning of many unnecessary nodes.}
    \label{fig:emergent_structure}
\end{figure*}

The implication of this is error correction converts many lethal mutations into \emph{neutral} mutations, such that organisms possessing error-correcting codes experience more neutral mutations than those without. This is widely recognized as a sign of an evolvable property \cite{Tenaillon2020, Wagner2008, Lenski2006}. It follows that error-correcting organisms can accumulate longer strings of mutations before suffering a lethal mutation, as shown in Fig.~\ref{fig:coding_exp}{b}. 

As illustrated in Fig.~\ref{fig:coding_exp}{c}, the lethal mutation fraction decreases with logical error probability. Therefore, having some error correction makes it easier to search for more error correction, which will lead to error correction being a typical result in noisy evolutionary processes. This is the essence of the principle of error correction enhanced evolvability. This principle explains the prevalence of error correction in our experimental results, and possibly the natural world. Interestingly, noise is required to bootstrap this process, as there is no evolutionary pressure to develop an initial, small error-correcting code without it. Noise enhances evolutionary processes. 

Mutations are unit distance movements in configuration space. Random sequential mutations trace out random walks in configuration space, and a lethal mutation occurs when this walk encounters a configuration with a fitness that is too low to be survivable. Therefore, these results may be equivalently understood as error correction locally smoothing over valleys in the fitness landscape being searched by evolution. Fig.~\ref{fig:coding_exp}{a} shows the smoothing of the region that can be reached by a single mutation and therefore may be interpreted as error correction reducing the magnitude of the "gradient" of the fitness with respect to the configuration (although this is not a true gradient, since the configuration space is discrete). Fig.~\ref{fig:coding_exp}{b} shows the same information over long distances.  This concept of error correction smoothing the local fitness landscape and allowing error-corrected organisms to search further for improvements is illustrated by the cartoon in Fig.~\ref{fig:intro_fig}.

Fig.~\ref{fig:emergent_structure} shows that our results extend to composite computational tasks. The most striking feature of these organisms is that when evolved in the presence of noise, the structure of the computation is strongly imprinted on the structure of the organism. This is shown clearly in Fig.~\ref{fig:emergent_structure}{a}. Here, a parallel AND task is used, in which two non-interacting AND computations must be completed. The task can be completed by two independent modules, each containing a pair of inputs and their corresponding outputs. When evolved in a noisy environment, this is exactly the structure present in the organisms. Independent modules saturate Newman's modularity metric \cite{newman_mod}: this solution could not be any more modular. When evolved in the absence of noise (with all other experimental parameters held constant) the structure vanishes, and the solutions resemble random graphs. The story is the same for tasks with more complicated modular structures, as shown in Fig.~\ref{fig:emergent_structure}{b} and c. It seems that noise, and noise alone, can be enough to drive the evolution of structural modularity in dynamic systems.

Although the idea of modularity emerging naturally to increase the robustness of systems to noise is old \cite{Wagner2007}, experimental demonstrations of this principle are rare. Understanding the emergence of modularity in cognition is a recent topic of interest. Most methods that have been developed involve imposing a special cost or environmental constraint that encourages modularity, such as rapidly varying a learning objective in a modular fashion \cite{Kashtan2001, Kashtan2007} and associating a cost with adding connections to a network \cite{Clune2013}. Results on noise alone leading to modularity are sparse, and limited to small neural networks \cite{hoverstad}. Therefore, our results on modularity taken alone are significant, as they demonstrate a long-believed concept within the physically relevant formalism of dynamic systems. 

It should be noted that like the primitive organisms, the organisms shown in Fig.~\ref{fig:emergent_structure} also possess error-correcting codes. As shown in Fig.~\ref{fig:emergent_structure}{d}, the structurally distinct modules seen in the shared AND organism also make up functionally distinct memory modules. Each memory module implements an independent maximum Hamming distance code for each output bit, akin to the AND code shown in Fig.~\ref{fig:intro_fig}. This modular error correction reflects the way the brain seems to implement independent error correcting codes in different regions \cite{Chaudhuri2019, Gardner2022, Sreenivasan2011}. The way the modules are constructed is also interesting, as from a fitness standpoint each output bit was made equally important and the same number of nodes are dedicated to each module. This kind of code is formally efficient in the sense that the number of bits it can encode scales linearly with the number of physical bits (it is asymptotically a constant rate code). The sequential AND organism develops a different kind of code, as shown in Fig.~\ref{fig:emergent_structure}{e}. In this case, memory modules would not be an efficient solution, as there are only 3 possible outputs for this task: 00, 01, and 11. Instead, this organism develops a distributed maximum Hamming distance code, reminiscent of Hamming codes themselves \cite{hamming1950}.

To explore the emergence of modularity at larger scales while maintaining computational tractability, instead of starting searches from random states one can seek to improve the fault tolerance of a network that already performs a particular computation in the absence of noise. In particular, we sample random networks that compute a function on a certain number of bits in the absence of noise, and then evolve them in a noisy environment and observe the results. An example of an organism that implements a six-bit function is shown in Fig.~\ref{fig:emergent_structure}{f}. We can see that substantial structure is evolved. The final network is modular, with nodes 1 and 3 being distinctly separated from the other nodes. Additionally, the evolutionary process removes many of the nodes present in the original network, suggesting they were useless to the computation and only increased noise cross-section. 

It can be hypothesized that the general principle behind noise-induced modularity is that noise punishes network size, both in terms of the number of connections and the number of nodes. Noise punishes connectivity because errors can propagate further and faster in more densely connected networks. For example, the separate modules evolved to solve the noisy parallel AND task prevent errors that strike one module from ever propagating to the other. In general, connectivity in the organisms evolved to solve noisy tasks reflects the computationally required connectivity to move information between inputs and outputs. The shared AND organisms generally separate into two modules, with connections present between the two only to communicate the state of node 1. Similar features are observed in the noisy sequential AND organisms. Noise punishes network size because larger networks have more components subject to error.

The experimental results presented in this paper demonstrate that error correction and modularity go hand-in-hand, and may be expected to emerge as a result of noisy evolutionary processes. Explicitly, we show that error correction is a typical result of evolving to perform a primitive computation in the presence of noise and that this principle scales to small composite computations, where solutions that are both modular and error correcting emerge. We reason that this occurs because of error correction enhanced evolvability: systems possessing error-correcting codes are easier to improve. With this principle in hand, we believe that these results should extend to larger scales more comparable with biological information processing systems. Such regimes are out of reach of our current techniques as evolving orders of magnitude larger organisms is presently computationally intractable. Future work may explore alternative modeling mechanisms, including direct physical implementations of noisy dynamics with electrical, mechanical, or biological mechanisms. 

\section*{Author Contributions}

TM designed and ran the experiments, and analyzed the results. TM, IRF, and ILC wrote and edited the manuscript. TM wrote the supplemental material.

\section*{Data and Code Availability}
The code used in this work to evaluate large batches of Boolean networks on GPUs has been open-sourced and is available on GitHub \cite{McCourt2023}. The code and data that can be used to reproduce the experiments and figures in the work are also available on GitHub \cite{papercode}.

~\\

\section*{acknowledgments}

This work was supported in part by the Institute for Artificial Intelligence and Fundamental Interactions (IAIFI) through NSF Grant No. PHY-2019786, and by NTT Research.
TM would like to thank Cian Schmitt-Ulms, Olive Garst, and Mikail Khona for their useful comments that helped clarify the manuscript.
TM would also like to thank the administrators of the subMIT cluster, on which many of the experiments presented in this work were completed. 

\bibliography{main}

\end{document}


\title{Supplementary material to "Noisy dynamical systems evolve error correcting codes and modularity"}
\author{Trevor McCourt, Ila R. Fiete, Isaac Chuang}
\date{March 22, 2023}
\maketitle

\section{Boolean Networks}\label{sec:net_spec}

\begin{figure}[H]
    \centering
    \includegraphics[width=\linewidth]{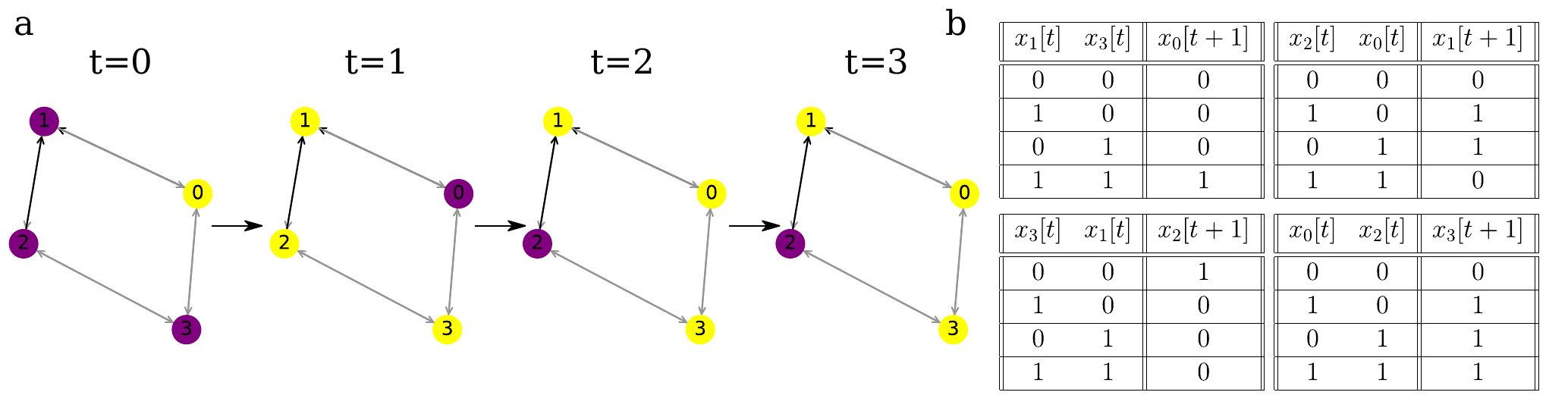}
    \caption{\textbf{An example Boolean network. a} A trajectory of an $N=4$ $k=2$ Boolean network. The network reaches a fixed point (steady-state) at $t=2$. \textbf{b} The truth tables associated with each of the $4$ nodes that, along with an initial state, generate the trajectory shown in (a).}
    \label{fig:net_fig}
\end{figure}

The state of an $N$-node Boolean network is given by a length $N$ binary vector $x$. The state of a particular node of a Boolean network $x_i$ is updated in time according to some discrete update rule,
\begin{equation}
    x_i[t+1] = f_i(x[t])
\end{equation}
Where $f_i$ is an arbitrary Boolean function. Traditionally, $f_i$ depends on $k \leq N$ of the $x$, such that each node in the network is influenced by exactly $k$ other nodes. Each $f_i$ can therefore be represented by a length $2^k$ truth table. Applying the update rule over and over again starting from some initial state generates a dynamic trajectory.  An example of an $N=4$ $k=2$ network is shown in Fig.~\ref{fig:net_fig}. 

This kind of constant $k$ Boolean network can be compactly specified by two matrices. The binary-valued function matrix $F$ has dimension $2^k \times N$, and has columns that are the output row of the truth table for each node. For example, for the network shown in Fig.~\ref{fig:net_fig},
\begin{equation}
    F = \begin{pmatrix}
0 & 0 & 1 & 0\\
0 & 1 & 0 & 1\\
0 & 1 & 0 & 1\\
1 & 0 & 0 & 1\\
\end{pmatrix}
\end{equation}
The second matrix is the integer-valued connectivity matrix $C$. This matrix has dimension $k \times N$, and the columns contain the node labels that influence each node. For the Fig.~\ref{fig:net_fig} network,
\begin{equation}
    C = \begin{pmatrix}
1 & 2 & 3 & 0\\
3 & 0 & 1 & 2\\
\end{pmatrix}
\end{equation}

\begin{figure}[H]
    \centering
    \includegraphics[width=0.5\linewidth]{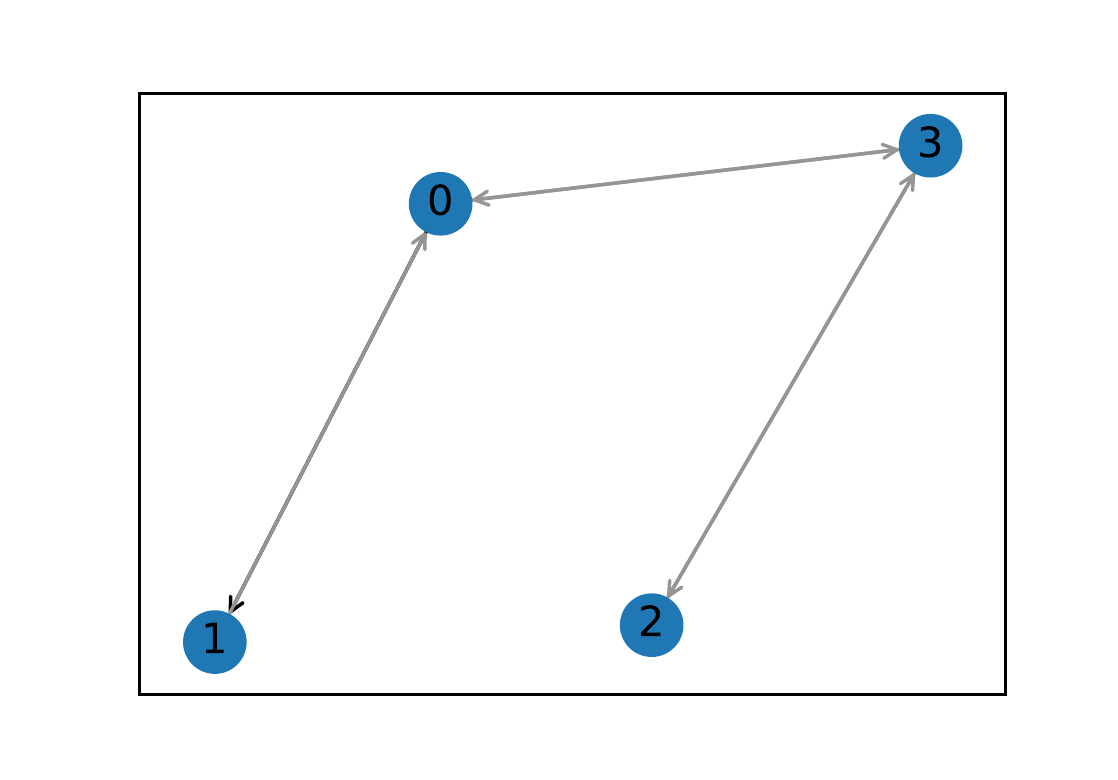}
    \caption{\textbf{An example ragged Boolean network} }
    \label{fig:rag_net_fig}
\end{figure}

We work with ragged networks, which are networks with variable $k$. Each node has its own degree $k_i \leq k_{max}$. An example of a ragged modification of the Fig.~\ref{fig:net_fig} network is shown in Fig.~\ref{fig:rag_net_fig}.

It is desirable to have a way of specifying ragged networks using rectangular data structures, such that batches of network specifications can be packed into tensors for accelerated simulation. To do this, we maintain the $F$ and $C$ matrices described above, which now are of dimension $2^{k_{max}} \times N$ and $k_{max} \times N$, respectively. We then define a binary masking matrix $U$ that indicates which connections in $C$ are active. For example, the $U$ matrix used to generate the Fig.~\ref{fig:rag_net_fig} network from the Fig.~\ref{fig:net_fig} network is,
\begin{equation}
    U = \begin{pmatrix}
1 & 0 & 1 & 1\\
1 & 1 & 0 & 1\\
\end{pmatrix}
\end{equation}
Therefore, a ragged boolean network can be specified by the triplet of matrices $(F, C, U)$.

\section{Accelerating Boolean Network Evaluation}

\begin{figure}[H]
    \centering
    \includegraphics[width=0.8\linewidth]{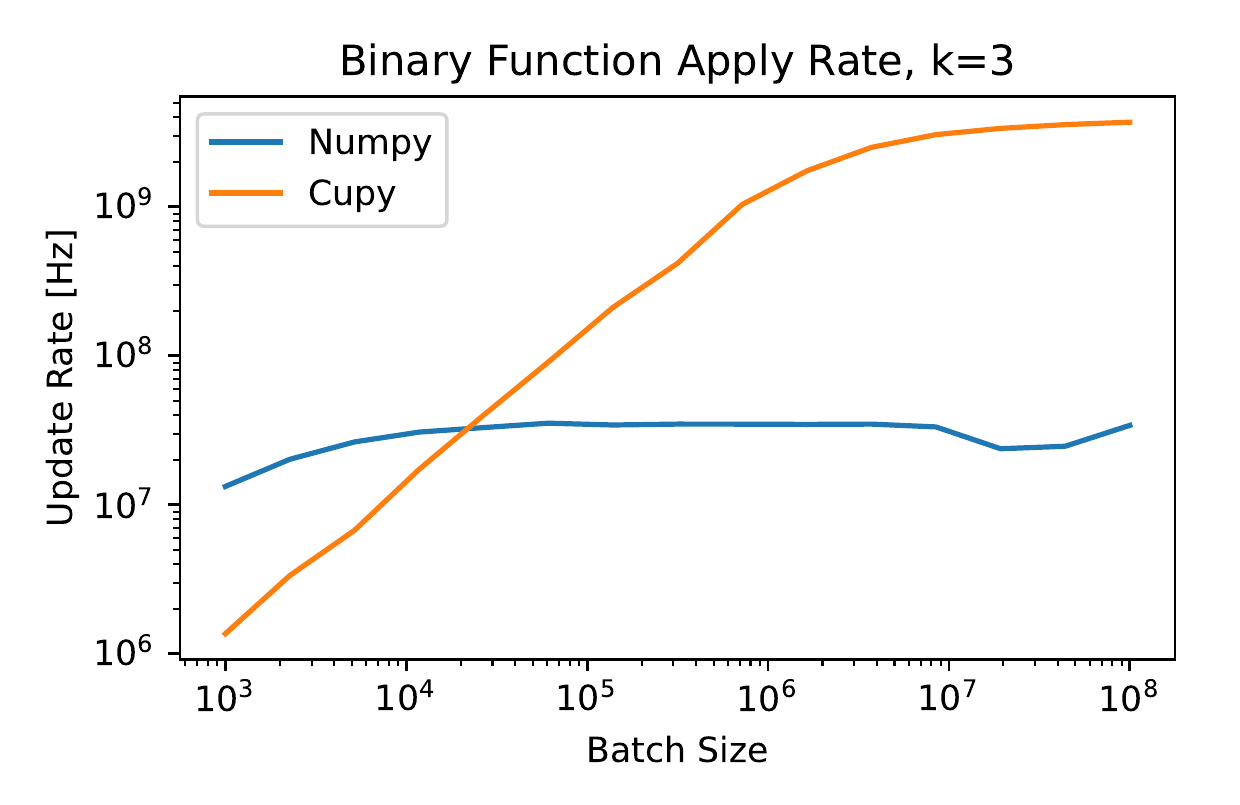}
    \caption{\textbf{Applying batches of binary functions on the CPU vs GPU} A GPU can accelerate the simulation of large batches of Boolean networks by more than 2 orders of magnitude. The GPU used here was an NVIDIA RTX 3090 and the CPU was an AMD Ryzen 9 5950x}
    \label{fig:accel_fig}
\end{figure}

With the data structures described in Sec.~\ref{sec:net_spec} in hand, it is straightforward to accelerate the evaluation of large batches of networks. The updating of a network can be handled using vectorized tensor operations. We've open-sourced our python implementation of the update rule on GitHub \cite{mccourt_2023}. Networks can be evaluated either on a CPU using the NumPy library \cite{harris2020array} or on an NVIDIA GPU using CuPy, \cite{cupy_learningsys2017}. In practice, GPUs substantially accelerate the evaluation of large batches of networks, see Fig.~\ref{fig:accel_fig}.

\section{XOR Evolution Details}

\begin{figure*}[h]
    \centering
    \includegraphics[width=0.8\linewidth]{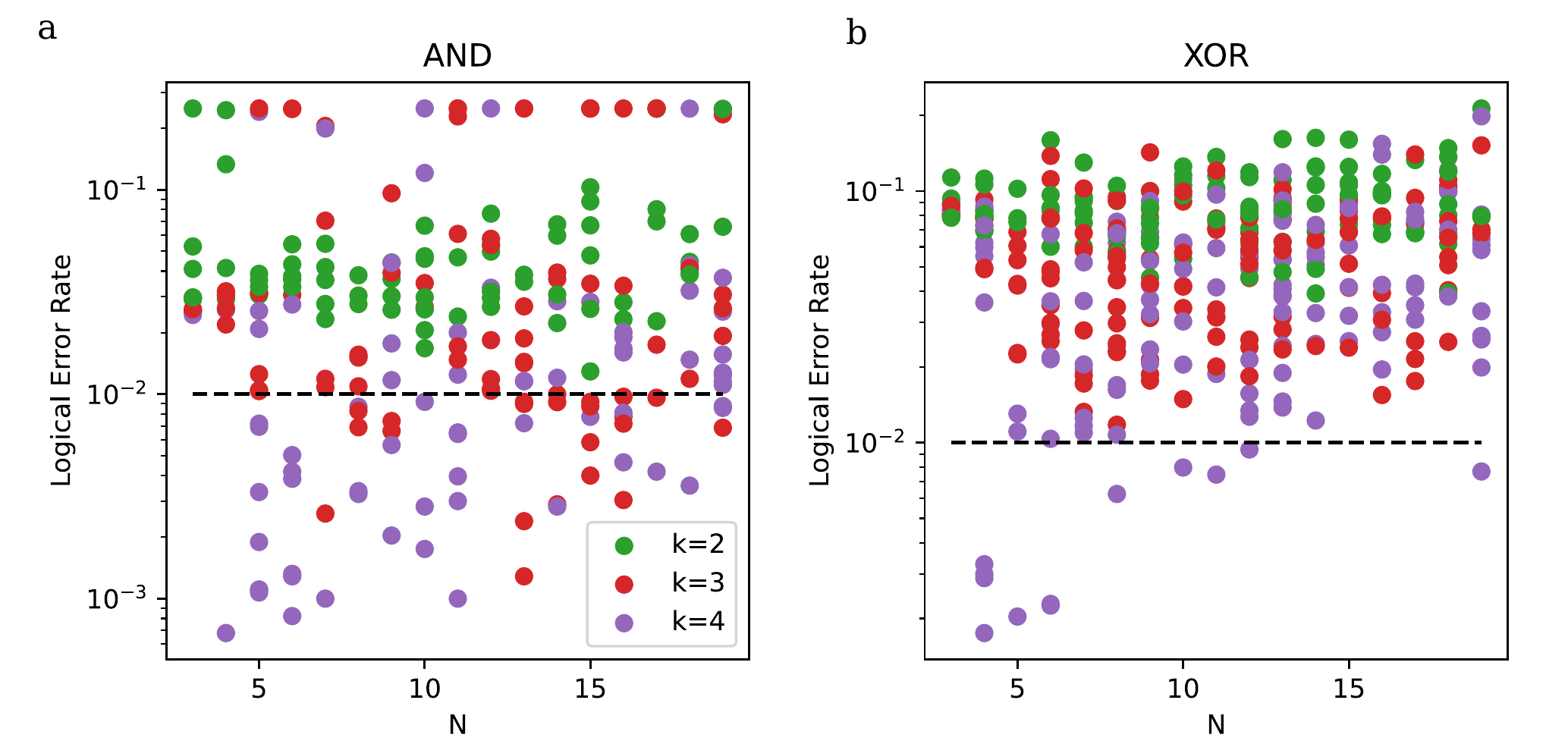}
    \caption{\textbf{The evolutionary characteristics of AND and XOR for different values of $N$ and $k_{max}$} Comparing the fitness of many independent AND (\textbf{a}) and XOR (\textbf{b}) populations with different values of $N$ and $k_{max}$ after 50k generations of evolution. It seems that $k=4$ is required for XOR to reach low error rates (at least in the same amount of time as AND). }
    \label{fig:and_vs_xor}
\end{figure*}

One curious difference between AND and XOR is that it seems higher connectivity is required to fault-tolerantly perform XOR. Fig.~\ref{fig:and_vs_xor} shows the results of many (short, only 50k generations run) evolutionary experiments comparing AND and XOR for many different values of $N$ and $k_{max}$.

This result is reasonable, as it is well known that XOR is a more complicated function to learn than AND \cite{minsky2017}. The reason for this is XOR has a more complicated truth table, with exactly half the bits being one (as opposed to only 1 bit for AND). As a result of this, we used $k_{max}=3$ for AND and $k_{max}=4$ for XOR in all of our experiments.

\begin{figure*}[h]
    \centering
    \includegraphics[width=\linewidth]{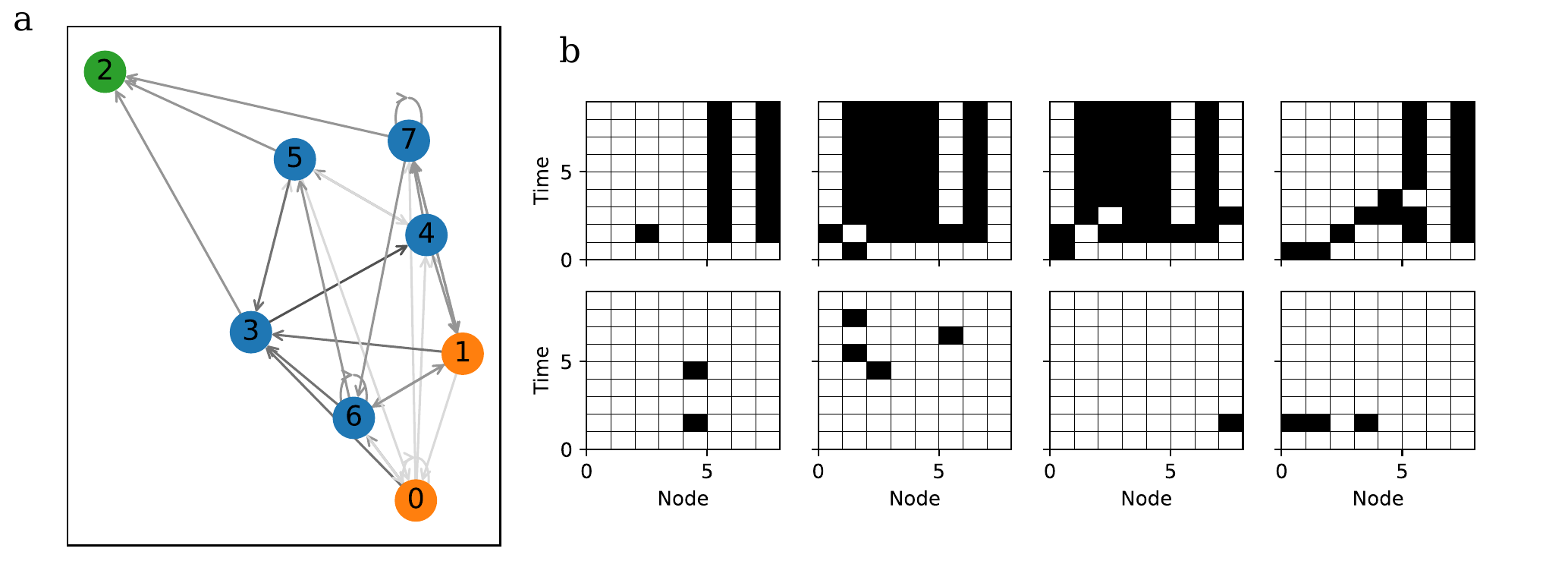}
    \caption{\textbf{An XOR network and a noisy trajectory a} A network that implements XOR. \textbf{b} A noisy trajectory of an XOR network.}
    \label{fig:xor_traj}
\end{figure*}

Fig.~\ref{fig:xor_traj} shows an example XOR network and noisy trajectory.

Note that the shown network only implements a distance 7 code, one bit is the same in each codeword. This seems to be typical of XOR networks, as shown in the main text figure 2. This is reasonable, as the evolutionary pressure to increase the code distance from 7 to 8 is low, and the optimization problems are more difficult for a $k_{max}=4$ network compared to a $k_{max}=3$ network, since the configuration space is larger and the loss landscape is more correlated \cite{kauffman1992}. 

\section{Effects of Various Crossover Operators}

\begin{figure}[H]
    \centering
    \includegraphics[width=0.8\linewidth]{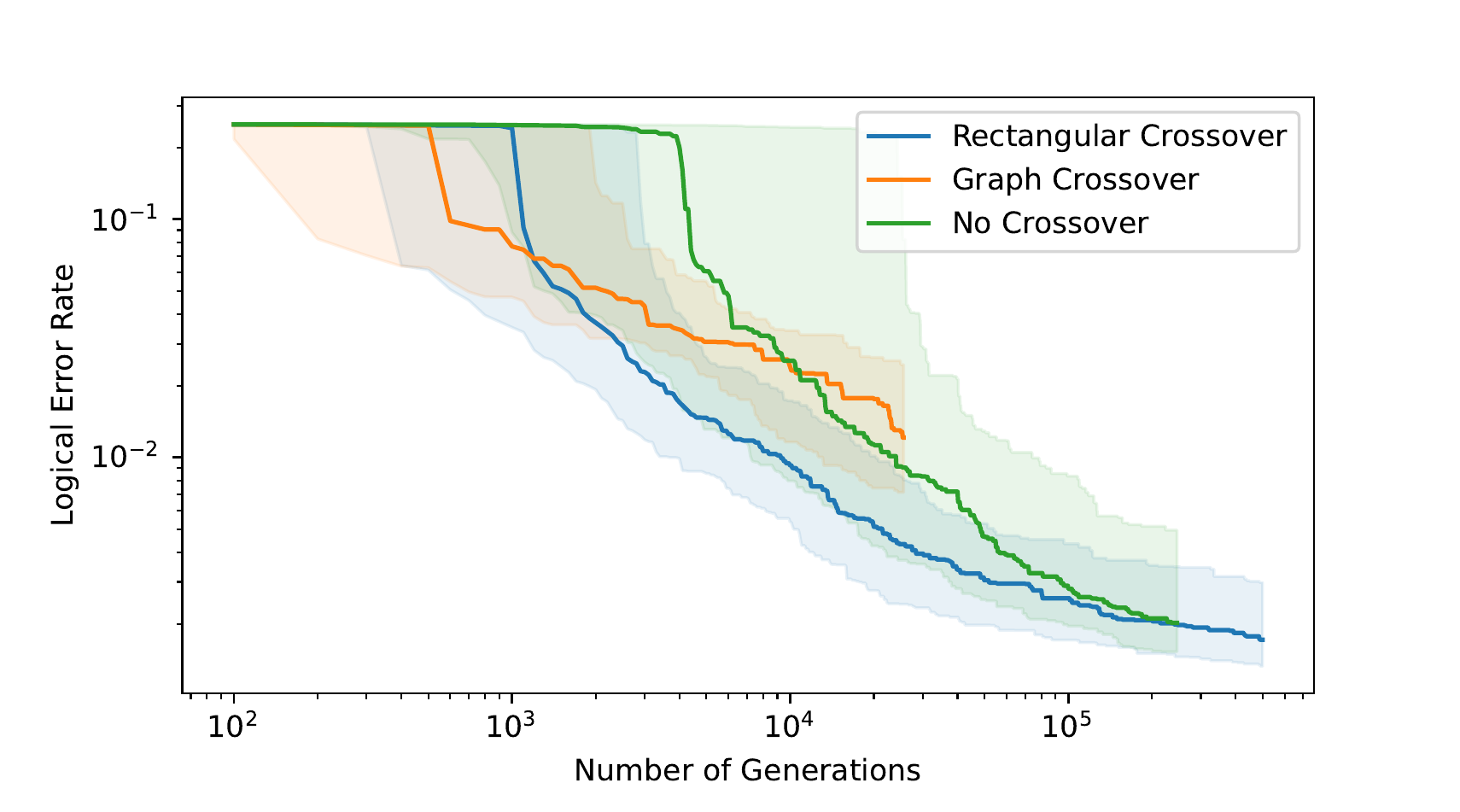}
    \caption{\textbf{Evolving AND with various crossover operators} }
    \label{fig:crossover_compare}
\end{figure}

Crossover operators are used in genetic algorithms to combine parent organisms to form a child organism, in the hope that the child will inherit the best features of all the parents and be of higher fitness. In a traditional genetic algorithm, crossover is applied to the fittest organisms in a population, and the children are then mutated and added back to the population. If no crossover operator is applied, we simply mutate one of the parents, and this process becomes equivalent to a pure random adaptive walk (greedily moving in random directions that improve fitness). We tried a few different crossover operators in this work and found that they were only slightly higher performing than a pure adaptive walk. This is shown in \ref{fig:crossover_compare}.

Rectangular crossover refers to splicing together the $F$, $C$, and $U$ matrices of two organisms. This is done by cutting each set of matrices vertically in the same position and re-assembling the components. This method does not respect the structure of the graphs, as the node label number is arbitrary. Graph crossover refers to cutting the graphs of two parents and wiring them back together in a way that respects the original connectivity \cite{Globus2013}.

The rectangular crossover slightly outperforms no crossover, converging to the same minimum error rate slightly before the random walk. This difference is small. Interestingly, it seems the graph crossover operation is the lowest performing, although it could not be tested to convergence as it is computationally intensive and requires transferring a large amount of data from GPU to CPU every generation. 

 \printbibliography